\documentclass[a4paper,12pt]{article}
\usepackage[cp1251]{inputenc}  
\usepackage[T2A]{fontenc}      
\usepackage[T2B]{fontenc}      
\usepackage[T2C]{fontenc}      
\usepackage[russian]{babel}    
\begin{document}
\begin{center}
CONSERVATION LAWS FOR CLASSICAL PARTICLES IN
ANTI--DE SITTER--BELTRAMI SPACE
\end{center}
\begin{center}
T. Angsachon, M. E. Chaikovskii, and  S. N. Manida

\end{center}

  In this paper we consider the conservation laws for classical particles in $AdS_4$.
   At first we parameterize a geodesic line and construct conserved quantities with analog
   of five dimensional Minkowski space-time $M^{(2,3)}$.

   Consequently we change $AdS_4$ space to AdS-Beltrami space-time and write conserved quantities
   in the Beltrami coordinates.
   Finally we take a limit for small velocity $\dot{x}\ll c$  and we get the conserved quantities
   in Lorentz-Fock space-time. And finally we give out the energy of the nonrelativistic noncosmological particle under the
   cosmological limit.
\section{Introduction}
It is known as the anti de-Sitter space-time ($AdS$) is the maximally symmetric solution of the Einstein field equation in the vacuum.
In this space-time there are 10 generators that equivalent to 10 generators of Minkowski space-time, but AdS space has the 2 universal constants.
These are the speed of light in vacuum c and curvature radius R which is respond to the cosmological constant$\lambda$. We only interest $AdS$ space-time in the Beltrami coordinates because in this coordinates the geodesic line is the straight line and we call them as $AdS$-Beltrami space-time.
In the first past we will define the AdS space-time and coordinates Beltrami. Furthermore we give out 10 generators of AdS-Beltrami space-time.
After these generators are find we will construct the conserved quantities in embeddind anti de-Sitter space-time($AdS_4$) and from this space
we also give out the 10 conserved quantities in AdS-Beltrami space-time. In the particularly we take the limit $R\rightarrow\infty$ for the consideration of nonrelativistic cosmological particles. When we take the cosmological limit ${(\lambda\dot{\vec{x}}-\vec{x})^2}\ll {R^2}$ we will receive the energy of nonrelativistic noncosmological particle which is different from usual galilean dynamics.
\section{Generators in AdS-Beltrami space-time}
We define $AdS$ space-time as five dimensional hyperboloid
\begin{equation}
AdS_4=\{ X \in M^{(2,3)} , X^2=\eta_{AB}X^A{}X^B=R^2 \}
\end{equation}
embedded in five dimensional Minkowski space-time. Indices A and B are running values -1,0,1,2,3.
Induced metric for this space can be expressed\\
\begin{equation}
ds^2=(\eta_{AB}dX^AdX^B)\mid_{AdS_4},
\end{equation}
where $\eta_{AB}=diag\{1,1,-1,-1,-1\}$.\\
This space-time has a corresponding isometry group $SO(2,3)$,  generated by the Killing vectors
\begin{equation}
L_{AB}=(X_A\frac{\partial}{\partial{X_B}}-X_B\frac{\partial}{\partial{X_A}}).
\end{equation}
We can define the generators which are corresponding with the Killing vectors
\begin{equation}
P_i=L_{-1i}, H=L_{0(-1)}, K_i=L_{i0}, J_i=\varepsilon_{ijk}L_{jk}
\end{equation}
where $i,j,k$ are running values 1,2,3.
The Beltrami coordinates can be defined as the projective coordinates on the half plane of hyperboloid in this formula
\begin{equation}
x_{\mu}=R\frac{X_\mu}{X_{-1}}
\end{equation}
We are only interested to find out the generators $P_i$ and H because of they have an analog of generators of 4-momentum and energy in
Minkowski space-time. The generators $K_i$ have the same form with boost generators and $J_i$ have the same form with generators of angular momentum
in Minkowski space-time.\\
For the construction of generator $P_i$ we consider the infinitesimal transformations between $X_i$ and $X_{-1}$ which are expressed that
\begin{equation}
X_{-1}'=X_{-1}\ch\alpha+X_i\sh\alpha \simeq X_{-1}+\alpha{}X_i
\end{equation}
\begin{equation}
X_i'=X_i\ch\alpha+X_1\sh\alpha \simeq X_i+\alpha{}X_{-1}
\end{equation}
where $\alpha$ is the infinitesimal parameter.
And we change these transformations to the form of Beltrami coordinates and give out generator in the formula
\begin{equation}
P_i = -\frac{x_ix_j}{R}\frac{\partial}{\partial x_j}+R\frac{\partial}{\partial x_i}-x_ix_0\frac{\partial}{\partial x_0}
\end{equation}
A generator $H$ also can find with the infinitesimal transformation between $X_0$ and $X_{-1}$
\begin{equation}
X_0' = X_0\cos\alpha+X_{-1}\sin\alpha \simeq X_0+\alpha X_{-1}
\end{equation}
\begin{equation}
X_{-1}' = -X_0\sin\alpha+X_{-1}\sin\alpha \simeq -\alpha X_0+X_{-1}
\end{equation}
After change these transformations into the Beltrami coordinates form the generator $H$ can be written out as the formula
\begin{equation}
H = (\frac{x_0^2}{R}+R)\frac{\partial}{\partial x_0}+\frac{x_0x_i}{R}\frac{\partial}{\partial x_i}
\end{equation}
A boost generator is received from the transformation between $X_0$ and $X_i$ which is the form in Minkowski space-time
\begin{equation}
K_i = x_0\frac{\partial}{\partial x_i}+x_i\frac{\partial}{\partial x_0}
\end{equation}
and generator of angular momentum also has the same form in Minkowski space-time
\begin{equation}
J_i = \varepsilon_{ijk}x_j\frac{\partial}{\partial x_k}
\end{equation}
These are 10 generators for the motion in AdS-Beltrami space-time.
\section{Conserved quantities in embedding anti de-Sitter space-time ($AdS_4$).}

For construction the action in this space we need choose initial points $X_0=(R,0,0,0,0)$, because of a group SO(3,2) transitively
actions on a manifold $AdS_4$.

Now we consider an action of the classical massive free particle, which can be expressed
in the five-dimensional coordinates
\begin{equation}
S=-mc\int[(V^2)^{\frac{1}{2}}+a(X^2-R^2)]d\lambda;
\end{equation}
where $X(\lambda)$ is a parameterized timelike curve with a constraint $X^2=R^2$.
$V^A(\lambda)=\frac{dX^A}{d\lambda}$ are the velocities of the particle. It's a tangent vector to
the curve $X(\lambda)$ and orthogonal to it ($X\cdot V=0$). We know that, an action (4 )is
invariant under transformation .
\begin{equation}
X_A \longrightarrow X_A+\omega_{AB}X^B
\end{equation}
where $\omega_{AB}$ is an antisymmetric infinitesimal parameter.

 From Noether theorem we can find
the conserved quantities along a timelike geodesic line
\begin{equation}
K_{AB}=\frac{m}{R\sqrt{V^2}}(X_AV_B-X_BV_A)
\end{equation}
Further we will describe conserved quantities in the form of parameterized timelike geodesic
line. Timelike geodesic line is an intersection between two dimensional plane, which contains
an origin of this space-time $M^{(2,3)}$ and $AdS_4$ hyperboloid.

We can choose  vectors $\xi$ and $\eta$, which are timelike ($\xi^2>0,\eta^2>0$) as,
\begin{equation}
\xi=(0,\xi_0,\vec{\xi}),\quad{}  \eta=(\eta_{-1},0,\vec{\eta})
\end{equation}
The geodesic line can be parameterized with the help of those vectors by parameter $\lambda$ as;
\begin{equation}
X=\frac{R(\lambda\xi+\eta)}{\sqrt{\lambda^2\xi^2+\eta^2+2\lambda(\xi\cdot\eta)}}.
\end{equation}
From this equation we can find a velocity
\begin{equation}
V=\frac{R}{\lambda^2\xi^2+\eta^2+2\lambda(\xi\cdot\eta)}
\left[\sqrt{\lambda^2\xi^2+\eta^2+2\lambda(\xi\cdot\eta)}\xi-\frac{(\lambda\xi+\eta)(\lambda\xi^2+\xi\cdot\eta)}
{\sqrt{\lambda^2\xi^2+\eta^2+2\lambda(\xi\cdot\eta)}}\right]
\end{equation}
We put (8) and (9) in the equation (6) and now conserved quantities are written in
a simple form
\begin{equation}
K_{AB}=\frac{m[\eta_A\xi_B-\eta_B\xi_A]}{\sqrt{\eta^2\xi^2-(\xi\cdot\eta)^2}}
\end{equation}
Those quantities are the components of the two-form:
\begin{equation}
K=K_{(\xi,\eta)}=m\frac{\eta\wedge\xi}{\sqrt{\eta^2\xi^2-(\xi\cdot\eta)^2}}
\end{equation}
Finally we can find a mass-shell equation
\begin{equation}
K_{AB}K^{AB}=2m^2
\end{equation}
\section{Conserved quantities in Beltrami coordinates.}
    In this part we will write the conserved quantities in the Beltrami coordinates.
Initially we choose the parameters $\xi=(0,c,\overrightarrow{\xi})$ and
 $\eta=(R,0,\overrightarrow{\eta})$, then we will write the equation of motion in
 four-dimensionally embedded space-time with formula
\begin{equation}
X_\mu=R\frac{\lambda(c,\overrightarrow{\xi})+(0,\overrightarrow{\eta})}
{\sqrt{\lambda^2\xi^2+\eta^2+2\lambda(\xi\cdot\eta)}}
\end{equation}\
and coordinate $X_{-1}$ is
\begin{equation}
X_{-1}=\frac{R^2}{\sqrt{\lambda^2\xi^2+\eta^2+2\lambda(\xi\cdot\eta)}}.
\end{equation}
Beltrami coordinates are determined form the relation (5)
\begin{equation}
x_{\mu}=R\frac{X_\mu}{X_{-1}}=(\lambda{}c,\lambda\overrightarrow{\xi}+\overrightarrow{\eta})
\end{equation}
We call the $AdS$ space-time in Beltrami coordinate that $AdS$--Beltrami space-time.
Let $H$,$P_i$,$K_i$,$J_i$ be conserved quantities in Beltrami coordinates, which  correspond
to conserved quantities in the embedded space-time\\
$H=K_{0(-1)}$,  $K_i=RK_{0i}$,  $J_i=Rc\varepsilon_{ijk}K_{jk}$,   $P_i=cK_{i(-1)}$.\\
Four dimensional space-time coordinates and its derivatives with respect to $\lambda$ are:
\begin{equation}
x_0=\lambda{}c,  x_i=\lambda\xi_i+\eta_i
\end{equation}
\begin{equation}
\frac{dx_0}{d\lambda}=\dot{x_0}=c, \frac{dx_i}{d\lambda}=\dot{x_i}=\xi_i
\end{equation}
Then a denominator in (10)is expressed as
\begin{equation}
\eta^2\xi^2-(\xi\cdot\eta)^2 = R^2c^2(1-\frac{\dot{\vec{x}}^2}{c^2}-
\frac{(\vec{x}-\lambda\dot{\vec{x}})^2}{R^2}+\frac{(\vec{x}\times\dot{\vec{x}})^2}{R^2c^2})
\end{equation}
Finally we can find the correspondence conserved quantities
\begin{equation}
H=K_{0(-1)}=\frac{m}{\sqrt{1-\frac{\dot{\vec{x}}^2}{c^2}-
\frac{(\vec{x}-\lambda\dot{\vec{x}})^2}{R^2}+\frac{(\vec{x}\times\dot{\vec{x}})^2}{R^2c^2}}}
\end{equation}
\begin{equation}
K_i=RK_{0i}=\frac{m(x_i-\lambda\dot{x_i})}{\sqrt{1-\frac{\dot{\vec{x}}^2}{c^2}-
\frac{(\vec{x}-\lambda\dot{\vec{x}})^2}{R^2}+\frac{(\vec{x}\times\dot{\vec{x}})^2}{R^2c^2}}}
\end{equation}
\begin{equation}
J_i=Rcm\frac{\varepsilon_{ijk}\dot{x_j}x_k}{\sqrt{1-\frac{\dot{\vec{x}}^2}{c^2}-
\frac{(\vec{x}-\lambda\dot{\vec{x}})^2}{R^2}+\frac{(\vec{x}\times\dot{\vec{x}})^2}{R^2c^2}}}
\end{equation}
\begin{equation}
P_i=cK_{i(-1)}=\frac{m\dot{x_i}}{\sqrt{1-\frac{\dot{\vec{x}}^2}{c^2}-
\frac{(\vec{x}-\lambda\dot{\vec{x}})^2}{R^2}+\frac{(\vec{x}\times\dot{\vec{x}})^2}{R^2c^2}}}
\end{equation}
And mass-shell equation in AdS-Beltrami is expressed in this formula
\begin{equation}
H^2+\frac{K_i^2}{R^2}+\frac{J_i^2}{R^2c^2}+\frac{P_i^2}{c^2} = 2m^2
\end{equation}
which have the same result with the mass-shell equation in embedding AdS space-time.
\section{Conservasion laws for nonrelativistic cosmological particles}
    We consider the conservation laws for particles which are moving with the small velocity
($\dot{x}_i\ll c$) in the space with constant curvature $R.$ There are the conserved quantities in
$AdS$-Beltrami space-time
\begin{equation}
H=\frac{m}{\sqrt{1-\frac{\dot{\vec{x}}^2}{c^2}-
\frac{(\lambda\dot{\vec{x}}-\vec{x})^2}{R^2}+\frac{(\vec{x}\times\dot{\vec{x}})^2}{R^2c^2}}}, \label{21}
\end{equation}
\begin{equation}
K_i=\frac{m(x_i-\lambda\dot{x_i})}{R\sqrt{1-\frac{\dot{\vec{x}}^2}{c^2}-
\frac{(\lambda\dot{\vec{x}}-\vec{x})^2}{R^2}+\frac{(\vec{x}\times\dot{\vec{x}})^2}{R^2c^2}}}, \label{22}
\end{equation}
\begin{equation}
P_i=\frac{m\dot{x_i}}{c\sqrt{1-\frac{\dot{\vec{x}}^2}{c^2}-
\frac{(\lambda\dot{\vec{x}}-\vec{x})^2}{R^2}+\frac{(\vec{x}\times\dot{\vec{x}})^2}{R^2c^2}}}. \label{23}
\end{equation}
In the limit $\frac{\dot{\vec{x}}^2}{c^2}\ll \frac{(\vec{x}-\lambda\dot{\vec{x}})^2}{R^2}<1$
the conserved quantities can be expressed as:
\begin{equation}
H=\frac{m}{\sqrt{1-\frac{(\lambda\dot{\vec{x}}-\vec{x})^2}{R^2}}},\label{121}
\end{equation}
\begin{equation}
K_i=\frac{m(x_i-\lambda\dot{x_i})}{R\sqrt{1-\frac{(\lambda\dot{\vec{x}}-\vec{x})^2}{R^2}}},\label{122}
\end{equation}
\begin{equation}
\pi_i\equiv \lim_{c\to\infty}cP_i=\frac{m\dot{x_i}}{\sqrt{1-
\frac{(\lambda\dot{\vec{x}}-\vec{x})^2}{R^2}}}. \label{123}
\end{equation}
Now the mass-shell equation contains only H and $K_i$:
\begin{equation}
H^2-K_i^2=m^2.
\end{equation}
The space-time with this mass-shell equation is called the Lorentz-Fock space-time.
If we put $\lambda=T+t, R/T\equiv c_0$ and consider the small vicinity point in this space-time
$t\ll T,$  $|\vec{x}|\ll R,$, then it is gived out
\begin{equation}
H=\frac{m}{\sqrt{1-\frac{\dot{\vec{x}}^2}{c_0^2}}},\label{321}
\end{equation}
\begin{equation}
\pi_i=\frac{m\dot{x_i}}{\sqrt{1-\frac{\dot{\vec{x}}^2}{c_0^2}}}. \label{323}
\end{equation}
\begin{equation}
K_i+\frac{1}{c_0}\pi_i=\frac{m(x_i-t\dot{x_i})}{\sqrt{1-\frac{\dot{\vec{x}}^2}{c_0^2}}},\label{322}
\end{equation}
Therefore we see that this "cosmological" dynamics under the nonrelativistic limit do not changes
from standard relativistic dynamics with speed of light $c_0$.

\section{Energy of nonrelativistic noncosmological particle}
    Now we change the expression for the energy (26) to the cosmological limit
${(\lambda\dot{\vec{x}}-\vec{x})^2}\ll {R^2}$ and write it in this form
\begin{equation}
H\simeq m+\frac{m(\vec{x}-\lambda\dot{\vec{x}})^2}{2R^2})+O(1/R^4)
\end{equation}
We put in the such as equations (30)-(32), $\lambda=T+\tau$, $\tau\ll T,$ $\frac{R}{T}\equiv c_0:$
\begin{equation}
\tilde{H}=Hc_o^2=mc_0^2+\frac{m\dot{\vec{x}}^2}{2}+c_0\frac{m\dot{\vec{x}}(t\dot{\vec{x}}-\vec{x})}{R}
+c_0^2\frac{m(t\dot{\vec{x}}-\vec{x})^2}{R^2}.
\end{equation}
The constructed quantity conserves without for any value $c_0$, and consequently
will be conserved in the expansion under any values of $c_0$:
\begin{equation}
\tilde{H_0}=\frac{m\dot{\vec{x}}^2}{2},
\end{equation}
\begin{equation}
\tilde{H_1}={m\dot{\vec{x}}(t\dot{\vec{x}}-\vec{x})},\label{h1}
\end{equation}
\begin{equation}
\tilde{H_2}=\frac{m(t\dot{\vec{x}}-\vec{x})^2}{2}.\label{h2}
\end{equation}
The conservation of the quantities (36),(37) is known as the symmetry consequence of Lagrangian
of the free point particles which is related with the symmetry of Schr\"{o}dinger group.
It is impossible to get these conservation laws from standard relativistic dynamics, but
we got them from cosmological dynamics in Lorentz-Fock space-time.
\section{Conclusion}
We have shown the conserved quantities in embedding $AdS$ and $AdS$-Beltrami space-time which have the 2 universal constants $c$ and $R$.
The conserved quantities in the another space-time can be given out by the limit change of the universal constant.
In Lorentz-Fock space-time we have get the mass-shell equation form the conserved quantities $H$ and $K_i$. And finally we have shown what
the new dynamics which is obtained by the cosmological time from energy in Lorentz-Fock space-time.
\begin{center}
\textbf{References}\\
\end{center}
1. S.Cacciatori,V.Gorini,A.Kamenshchik, Special Relativity in the $21^{st}$ century,
hep-th/0807.3009.\\
2. S. Cacciatori, V. Gorini, A. Kamenshchik and U. Moschella, Conservation laws and scattering for de Sitter classical particles,
Class.Quant.Grav.25:075008,2008.\\
3. Han-Ying Guo, Special Relativity and Theory of Gravity via Maximum Symmetry and Localization, Science in China A, Vol.51 [4] (2008) 568-603,
gr-qc/0707.385.\\
4. S.N. Manida, Fock-Lorentz transformations and time-varying speed of light,
gr-qc/9905046.
\end{document}